\documentclass[useAMS,twocolumn,usenatbib]{mn2e}

\usepackage{graphicx}
\usepackage{amssymb,amsmath}

\begin{document}
%%%%%%%%%%%%%%%%%%%%%%%%%%%%%%%%%%%%%%%%%%%%%%%%%%%%%%%%%%%%%%%%%%%%%%%

\title[Shaping the galaxy stellar mass function with winds]
{Shaping the galaxy stellar mass function with supernova- and AGN-driven winds}
\author[Ewald Puchwein \& Volker Springel]
{Ewald Puchwein$^{1}$ and
Volker Springel$^{1,2}$
\\$^1$Heidelberger Institut f{\"u}r Theoretische Studien,
Schloss-Wolfsbrunnenweg 35,
69118 Heidelberg, Germany
\\$^2$Zentrum f\"ur Astronomie der Universit\"at Heidelberg, Astronomisches Recheninstitut, M\"{o}nchhofstr. 12-14, 69120 Heidelberg, Germany}
\date{\today}
\maketitle

\begin{abstract} 
  Cosmological hydrodynamical simulations of galaxy formation in
  representative regions of the Universe typically need to resort to
  subresolution models to follow some of the feedback processes
  crucial for galaxy formation. Here, we show that an energy-driven
  outflow model in which the wind velocity decreases and the wind mass
  loading increases in low-mass galaxies, as suggested by
  observations, can produce a good match to the low-mass end of the
  observed galaxy stellar mass function. The high-mass end can be
  recovered simultaneously if feedback from active galactic nuclei
  (AGN) and a correction for diffuse stellar light plausibly missed in
  observations are included. At the same time, our model is in good
  agreement with the stellar mass functions at redshifts $z=1$ and
  $z=2$, and with the observed redshift evolution of the cosmic star
  formation rate density. In addition, it accurately reproduces the
  observed gas to stellar mass ratios and specific star formation
  rates of galaxies as a function of their stellar mass. This
  agreement with a diverse set of data marks significant progress in
  hydrodynamically modelling the formation of a representative galaxy
  population.  It also suggests that the mass flux in real galactic
  winds should strongly increase towards low-mass galaxies. Without
  this assumption, an overproduction of galaxies at the faint-end of
  the galaxy luminosity function seems inevitable in our
  models.  \end{abstract}


%%%%%%%%%%%%%%%%%%%%%%%%%%%%%%%%%%%%%%%%%%%%%%%%%
%%%%%%%%%%%%%%%%%%%%%%%%%%%%%%%%%%%%%%%%%%%%%%%%%
\section{Introduction}
\label{sec:introduction}
%%%%%%%%%%%%%%%%%%%%%%%%%%%%%%%%%%%%%%%%%%%%%%%%%
%%%%%%%%%%%%%%%%%%%%%%%%%%%%%%%%%%%%%%%%%%%%%%%%%

The formation and evolution of galaxies is regulated by a complex
interplay of accretion, star formation, and associated feedback
processes.  Unravelling the detailed physics involved is one of the
greatest challenges in modern astrophysics. High-resolution
cosmological simulations guided by increasingly accurate observations
have become one of the most important tools for tackling this problem
\citep[e.g.][]{Jenkins1998, Springel2005a, Boylan-Kolchin2009,
  Klypin2011}. However, even with today's fastest supercomputers, it
is numerically still infeasible to fully resolve all the relevant
processes that govern the formation of stars in galaxies and their
back-reaction on the interstellar medium (ISM), even though remarkable
progress towards this goal has recently been made for simulations of
isolated galaxies \citep{Hopkins2011, Hopkins2012b,
  Guedes2011}. Thus, hydrodynamical simulations of full galaxy
populations \citep[e.g.][]{Crain2009, Schaye2010, Vogelsberger2011}
usually employ simplified sub-resolution models to capture star
formation and associated feedback processes in an approximate way,
with parameterizations that are guided by observations. So-called
semi-analytical galaxy formation models \citep[e.g.][]{Guo2011} go yet
a step further and also replace the hydrodynamics itself with a highly
approximate schematic treatment, using as input halo merger trees
obtained from collisionless simulations.

An important difficulty that every theoretical model of galaxy
formation faces within the $\Lambda$CDM cosmogony is to explain the
very different shape of the dark matter halo mass function and the
galaxy stellar mass function (GSMF). The halo mass function exhibits
an almost featureless power-law with a steep logarithmic slope of $d
\log n / d \log M \sim -0.9$ for halo masses below $\sim 10^{13}
h^{-1} M_\odot$ \citep[e.g.,][]{Tinker2008}. The GSMF, instead, shows
a relatively flat slope at the low mass end, $d \log \phi / d \log M
\sim -0.2$, and a break (known as the `knee') at a characteristic
stellar mass of $\sim 5 \times 10^{10} h^{-1} M_\odot$, with an
exponential suppression of the abundance of still larger galaxies
\citep[e.g.,][]{Li2009}.

Cosmological hydrodynamical simulations that account for radiative
cooling but lack feedback processes associated with star formation
fail to reproduce this behaviour and strongly overpredict the stellar
mass fractions of halos \citep{Balogh2001}. It has hence been realized
early on that some kind of feedback mechanism needs to be included
\citep{White1991}. However, as the relevant physical scales of the
multiphase ISM or of AGN can not be resolved in large volume
simulations, previous work has introduced a number of phenomenological
sub-resolution models for treating such feedback effects
\citep{Springel2003, Schaye2008, Springel2005b, Borgani2005,
  Sijacki2007, Booth2009}. In most works, the primary feedback physics
invoked has been supernovae, which are thought to be crucial for
reducing the cosmic star formation rate to reasonable levels. However,
different approaches were used in numerical models to account for the
feedback energy supplied by supernova explosions. Many implementations
inject at least part of the energy thermally into the ISM to
counteract radiative cooling losses.  However, in its simplest form,
such an approach is relatively ineffective and does not create
outflows, prompting a number of `delayed cooling' models
\citep[e.g.][]{Thacker2000, Scannapieco2006} that try to suppress a
rapid cooling loss of the injected heat energy. Alternatively, a
kinetic feedback component which directly drives galactic winds has
often been invoked \citep[][]{Springel2003}.

In the simplest parameterization, winds are
driven with a fixed mass loading $\eta$, i.e.~by assuming a fixed
ratio between wind mass flux and the galaxy's star formation rate. If
the assumed amount of energy available for the kinetic feedback per
unit mass of stars formed \citep{Springel2003} is also kept fixed, a
constant wind velocity results. Hence we will call such models {\it
  constant wind} models. They were found to significantly reduce the
abundance of galaxies with stellar masses below a few times $10^{10}
M_\odot$, but the strength and exact mass-scale of the suppression
depends on the detailed choice of the wind parameters \citep[e.g.][see
also our Fig.~\ref{fig:faint_cw}]{Crain2009}. These simulations have,
however, so far neither been able to reproduce the slope of the
low-mass end of the observed GSMF nor its overall shape and
normalization. Furthermore, simulations that did not include AGN
feedback tended to overpredict the abundance of massive galaxies.

In comparison, semi-analytical models have been much more successful
in reproducing the global properties of galaxies, including their
abundance as a function of stellar mass (i.e. the GSMF) and the
redshift evolution of primary characteristics of the galaxy population
\citep[e.g.,][]{Croton2006, Bower2006}. From such studies it was
concluded that the faint-end slope of the galaxy luminosity function
can be made compatible with observations if the ratio of the mass flux
of gas ejected from galactic discs by supernova and stellar winds to
the star formation rate is scaled with properties of the host galaxy.
This has often been parameterized by scaling the ejection efficiency
with an inverse power of the disc rotation velocity, $\sim v_{\rm
  disc}^{-\alpha}$, with typical values for $\alpha$ in the range 2 to 3.2
\citep{Cole2000, Benson2003, Bower2012}. It was also found that by
including simple models for AGN feedback the abundance of massive
galaxies could be significantly reduced and made compatible with
observations \citep[e.g.,][]{Croton2006, Bower2006}.

Observationally, the mass loadings and velocities of galactic winds
are still not accurately constrained, but some progress has been made
in recent years by measuring Na, Na {\sc i} and Mg {\sc ii}
absorption lines in outflows. \citet{Martin2005} discovered a
correlation between the wind velocities and star formation rates of
galaxies, $v_{\rm w} \sim {\rm SFR}^{0.35}$. For galaxies with
measured rotation curves, a positive correlation between the galaxies'
circular velocities and the terminal wind velocities was 
found, and maximum wind velocities slightly smaller than the galaxies'
escape velocities were inferred. \citet{Martin2006} estimated mass
outflow rates in the winds' {\it cold} components of the order of 10\%
of the SFR. The achieved level of accuracy did, however, not allow
firm conclusions about how the mass loading scales with wind
velocity. \citet{Weiner2009} observed a similar correlation between
SFR and outflow velocity, $v_{\rm w} \sim {\rm SFR}^{0.3}$, in $z \sim
1.4$ galaxies, as well as maximum wind velocities close to the escape
velocity. \citet{Chen2010} found weak (yet noisy) positive
correlations between outflow velocity and star formation rate density
or galaxy stellar mass, respectively.

Motivated by these observations and the successes of semi-analytical
models, first attempts have been made to include phenomenological wind
models in cosmological hydrodynamical simulations in which the
wind velocity and mass loading scale with galaxy properties. We
will refer to such models as {\it variable wind}
models. \citet{Oppenheimer2006} introduced a scaling of wind
properties with galaxy velocity dispersion $\sigma$ by using the
gravitational potential $\phi$ at the location of a galaxy as a proxy
for $\sigma$, based on the approximation $\sigma \sim \sqrt{-1/2
  \phi}$. Assuming that winds are momentum driven \citep[see,
e.g.,][]{Murray2005}, they then adjusted the wind velocity $v_{\rm w}$
and wind mass-loading $\eta$ in their simulations according to $v_{\rm
  w} \sim \sigma$ and $\eta \sim 1/v_{\rm w} \sim 1/\sigma$. Compared
to previous wind models this resulted in more realistic metallicities
of the IGM as constrained by C {\sc iv} absorption studies. In
addition, the observed cosmic star formation rate density at high
redshift is reproduced better \citep[see also][]{Schaye2010}.

\citet{Okamoto2010} assumed an energy-driven wind scenario in which the wind properties scale according to $v_{\rm w} \sim \sigma$ and $\eta \sim 1/v_{\rm w}^2 \sim 1/\sigma^2$. Their model predictions were shown to be in broad agreement with the observed luminosity function of Milky Way satellites. \citet{Dave2011a} used an updated version of the \citet{Oppenheimer2006} momentum-driven model in which the galaxy velocity dispersion is instead estimated based on halo masses computed by an on-the-fly friend-of-friend group finder \citep{Oppenheimer2008}. They demonstrated that this model can roughly reproduce the faint-end of the GSMF \citep[also see][]{Oppenheimer2010, Choi2011} and the specific star formation rates of galaxies in the $10^{10} - 10^{11} M_\odot$ stellar mass range. Furthermore, their variable wind simulations match the slope of the gas-phase metallicity -- stellar mass relation better, and produce more realistic gas mass to stellar mass ratios in massive galaxies \citep{Dave2011b}.

Overall, variable winds are clearly more successful in suppressing star formation in low mass galaxies due to the boosted wind mass loading in these objects, as well as in predicting more realistic ISM masses and metallicities. In this work, we will hence explore variable wind models further. We will outline our numerical methods and describe our simulation set in Section~\ref{sec:methods}, discuss how the $z=0$ GSMF depends on the details of the wind model in Section~\ref{sec:gsmf_faint}, and investigate the impact of AGN feedback and the influence of diffuse light on the abundance of massive galaxies in Section~\ref{sec:gsmf_bright}. In Section~\ref{sec:sfh}, we assess whether the wind model that reproduces the $z=0$ GSMF best can also match the cosmic star formation history, as well as the GSMFs at redshifts $z=1$ and $2$. We finally explore the impact of winds on the gas masses and specific star formation rates of galaxies in Section~\ref{sec:fgas}, and summarize our findings in Section~\ref{sec:conclusions}.

%%%%%%%%%%%%%%%%%%%%%%%%%%%%%%%%%%%%%%%%%%%%%%%%%
%%%%%%%%%%%%%%%%%%%%%%%%%%%%%%%%%%%%%%%%%%%%%%%%%
\section{Methodology}
\label{sec:methods}
%%%%%%%%%%%%%%%%%%%%%%%%%%%%%%%%%%%%%%%%%%%%%%%%%
%%%%%%%%%%%%%%%%%%%%%%%%%%%%%%%%%%%%%%%%%%%%%%%%%

%%%%%%%%%%%%%%%%%%%%%%%%%%%%%%%%%%%%%%%%%%%%%%%%%
\subsection{The simulations}
\label{sec:simulations}
%%%%%%%%%%%%%%%%%%%%%%%%%%%%%%%%%%%%%%%%%%%%%%%%%

Throughout this paper, we make use of a set of cosmological hydrodynamical simulations that were performed with the TreePM-SPH simulation code {\sc P-Gadget-3}, an updated and significantly extended version of {\sc Gadget-2} \citep{Springel2005c}. The simulations adopt the best-fit WMAP 7-year cosmology \citep{Komatsu2011} with $\Omega_{\rm M}=0.272$, $\Omega_{\Lambda} = 0.728$, $\Omega_{\rm B}=0.0465$, $h=0.704$, and $\sigma_{8}=0.809$.

In this work, we want to explore the simulated galaxy populations and galaxy stellar mass functions both at the low-mass and the high-mass end. This calls simultaneously for a high mass resolution (to resolve the small objects) and a large volume (to find the rare high mass objects). An immense computational effort would be necessary to meet these requirements in a single simulation. To avoid this problem, we have chosen to perform a set of simulations that cover a range of box sizes from $15$ to $60\, h^{-1} \mathrm{Mpc}$ and a factor of 64 in mass resolution. The parameters of all runs that we use in this work are summarized in Table~\ref{tab:sims}.

%================================================
\begin{table*}
\begin{tabular}{lcccccccccc}
\hline
simulation model name & box size & $n_{\rm part}$ & $m_{\rm gas}$ & $\epsilon$ & $v_{\rm w}$ & $\eta$ & $f_{\rm UV}$ & IMF \\
& $[\,h^{-1} \mathrm{Mpc}]$ & & $[\,h^{-1} M_\odot]$ & $[\,h^{-1} \mathrm{kpc}]$ & & & &\\
\hline
L15N128csf & 15 & $2\times128^3$ & $2.0\times10^7$ & 4.5 & - & - & 1 & Sal.\\
L15N128csf,cw & 15 & $2\times128^3$ & $2.0\times10^7$ & 4.5 & $484 \, {\rm km} \, {\rm s}^{-1}$ & 2 & 1 & Sal. \\
L15N128csf,cw,polar & 15 & $2\times128^3$ & $2.0\times10^7$ & 4.5 & $484 \, {\rm km} \, {\rm s}^{-1}$ & 2 & 1 & Sal. \\
L15N128csf,mdvw1.25 & 15 & $2\times128^3$ & $2.0\times10^7$ & 4.5 & $1.25 \times v_{\rm esc}$ & $\sim v_{\rm w}^{-1}$ & 1 & Sal. \\
L15N128csf,edvw1.25 & 15 & $2\times128^3$ & $2.0\times10^7$ & 4.5 & $1.25 \times v_{\rm esc}$ & $\sim v_{\rm w}^{-2}$ & 1 & Sal.\\
L15N128csf,edvw0.6 & 15 & $2\times128^3$ & $2.0\times10^7$ & 4.5 & $1.25 \times v_{\rm esc}$ & $\sim v_{\rm w}^{-2}$ & 1 & Sal.\\
L15N128csf,edvw0.6,uv,chab & 15 & $2\times128^3$ & $2.0\times10^7$ & 4.5 & $0.6 \times v_{\rm esc}$ & $\sim v_{\rm w}^{-2}$ & 3 & Chab.\\
L15N256csf & 15 & $2\times256^3$ & $2.5\times10^6$ & 2.25 & - & - & 1 & Sal.\\
L15N256csf,cw & 15 & $2\times256^3$ & $2.5\times10^6$ & 2.25 & $484 \, {\rm km} \, {\rm s}^{-1}$ & 2 & 1 & Sal. \\
L15N256csf,edvw0.6,uv,chab & 15 & $2\times256^3$ & $2.5\times10^6$ & 2.25 & $0.6 \times v_{\rm esc}$ & $\sim v_{\rm w}^{-2}$ & 3 & Chab.\\
L15N512csf,edvw0.6,uv,chab & 15 & $2\times512^3$ & $3.2\times10^5$ & 1.125 & $0.6 \times v_{\rm esc}$ & $\sim v_{\rm w}^{-2}$ & 3 & Chab.\\
L60N512csf & 60 & $2\times512^3$ & $2.0\times10^7$ & 4.5 & - & - & 1 & Sal.\\
L60N512csf,cw & 60 & $2\times512^3$ & $2.0\times10^7$ & 4.5 & $484 \, {\rm km} \, {\rm s}^{-1}$ & 2 & 1 & Sal.\\
L60N512csf,edvw0.6,uv,chab & 60 & $2\times512^3$ & $2.0\times10^7$ & 4.5 & $0.6 \times v_{\rm esc}$ & $\sim v_{\rm w}^{-2}$ & 3 & Chab.\\
L60N512csfbh,edvw0.6,uv,chab & 60 & $2\times512^3$ & $2.0\times10^7$ & 4.5 & $0.6 \times v_{\rm esc}$ & $\sim v_{\rm w}^{-2}$ & 3 & Chab.\\
\hline
\end{tabular}
\caption{Summary of the parameters of the simulations used in this work. Given are the comoving box size, the number of simulation particles $n_{\rm part}$ in the initial conditions (half of them are gas and half of them are dark matter particles), the gas particle mass $m_{\rm gas}$, the comoving gravitational softening (Plummer equivalent), the wind velocity $v_{\rm w}$, the wind mass loading $\eta$, the UV heating boost factor $f_{\rm UV}$, and the assumed IMF (Salpeter or Chabrier).}
\label{tab:sims}
\end{table*}
%================================================

Strictly speaking, it is problematic to run simulations with a comoving box size of only $15\, h^{-1} \mathrm{Mpc}$ until redshift $z=0$. At low redshift the fundamental mode in such small simulation boxes becomes non-linear, but since a coupling to larger modes is absent, the evolution of these modes will not be fully correct. The simulation box may then not be a representative sample of the universe any more. We checked, however, that the quantities in which we are most interested in, i.e. the GSMF and the stellar mass fraction of halos, are not significantly affected by finite box-size effects other than small-number statistics at the high-mass end (compare, e.g., the {\it solid green} curves in Figs.~\ref{fig:faint_vel_uv_imf} and \ref{fig:bright}). The $15\, h^{-1} \mathrm{Mpc}$ boxes can thus be reliably used to investigate the effects of winds on the low-mass end. For the high-mass end, we use simulations with a box size of $60\, h^{-1} \mathrm{Mpc}$.

%%%%%%%%%%%%%%%%%%%%%%%%%%%%%%%%%%%%%%%%%%%%%%%%%
\subsection{The wind model}
\label{sec:wind_model}
%%%%%%%%%%%%%%%%%%%%%%%%%%%%%%%%%%%%%%%%%%%%%%%%%

Inspired by the observations discussed in Section~\ref{sec:introduction}, we want to study how galaxy formation is affected by galactic winds whose velocities and mass loadings scale with the properties of the galaxies form which they are launched. More specifically, in our simulations we assume that the wind velocity $v_{\rm w}$ is directly proportional to a galaxies' escape velocity $v_{\rm esc}$, i.e.  \begin{equation}
  v_{\rm w} = \kappa \times v_{\rm esc}.
\label{eq:wind_vel}
\end{equation}
The choice of the proportionality constant $\kappa$ will be discussed in more detail in Section~\ref{sec:gsmf_faint}. The escape velocity from the centre of a galaxy that is well described by a Navarro-Frenk-White (NFW) profile \citep{Navarro1996} is given by
\begin{equation}
  v_{\rm esc} = v_{200 \rm c} \times \sqrt{\frac{2 \, c}{\ln(1+c) - \frac{c}{1+c}}},
\label{eq:vesc}
\end{equation}
where $c \equiv r_{\rm s}/r_{200 \rm c}$ is the halo's concentration, i.e.~the ratio of the NFW-scale radius $r_{\rm s}$ to the radius $r_{200 \rm c}$ at which the mean enclosed density is 200 times the critical density of the Universe. $v_{200 \rm c} = \sqrt{G M_{200 \rm c} / r_{200 \rm c}}$ is the circular velocity at $r_{200 \rm c}$, where $G$ is the gravitational constant, and $M_{200 \rm c}$ is the mass enclosed within radius $r_{200 \rm c}$. In our simulations, we run an on-the-fly friends-of-friends (FoF) group finder at regular intervals and store 
for each gas particle the mass of the FoF-group it is part of. Based on this mass, the host halo's escape velocity can be estimated using Eq.~(\ref{eq:vesc}) and the mass-concentration relation $c=4.67 \times (M_{200 \rm c} / [10^{14} h^{-1} M_\odot])^{-0.11}$ obtained by \citet{Neto2007} from the Millennium N-body simulation \citep{Springel2005a}.\footnote{As the wind velocity is not very sensitive to the mass-concentration relation, we ignore redshift evolution effects and use the $z=0$ relation from \citet{Neto2007} also at $z>0$.}

Once we have fixed the velocities of galactic winds, we need a prescription for the amount of gas that enters into such winds. A simple choice that was used in most previous sub-resolution models of wind formation is to link the power that is available for driving the wind to the star formation rate of the galaxy from which it is launched. In particular, the wind energy has often been assumed to be a specific fraction of the available supernova feedback energy \citep[see e.g.][]{Springel2003}. The latter is proportional to the star formation rate, with a proportionality constant that depends on the assumed initial stellar mass function (IMF). Under these assumptions, the mass flux $\dot{M}_{\rm w}$ of the wind is related to the galaxy's star formation rate $\dot{M}_{*}$ by
\begin{equation}
 \frac 12 \dot{M}_{\rm w} v^2_{\rm w} = \epsilon_{\rm SN,w} \dot{M}_{*},
\end{equation}
where $\epsilon_{\rm SN,w}$ is the energy available to the wind per unit mass of long-lived stars formed. For such {\it energy-driven} winds, the mass-loading $\eta \equiv  \dot{M}_{\rm w} / \dot{M}_{*}$ scales with $\eta \sim v^{-2}_{\rm w}$. Thus, winds with velocities given by Eq. (\ref{eq:wind_vel}) will have significantly larger mass-loadings in low-mass halos and are therefore expected to suppress star formation more strongly there.

In reality, the amount of energy that is available for driving a galactic wind may not only depend on the galaxy's star formation rate. For example, the fraction of injected energy that is lost by radiative cooling could depend on halo mass. On the other hand, momentum that is injected by stellar winds, supernovae and AGN cannot be radiated away. Accordingly, if galactic winds are {\it momentum-driven}, it would seem more natural to assume a proportionality between the winds' momentum flux (rather than its energy flux) and the star formation rate of the host galaxy. In this case, the mass-loading would scale like $\eta \sim v^{-1}_{\rm w}$. 

In this work, we will explore the effects of both a $\sim v^{-2}_{\rm w}$ and a $\sim v^{-1}_{\rm w}$ dependence of the mass-loading on the wind velocity, where we assume in both cases that the wind velocity is related to the halo's escape velocity by Eq. (\ref{eq:wind_vel}). For reference, we will also show results based on the galactic wind model originally introduced by \citet{Springel2003}, which assumes a constant wind mass-loading $\eta$ and wind velocity $v_{\rm w}$.

Unless mentioned otherwise, we assume our winds to be isotropic, i.e.~gas particles entering the wind near the centre of a galaxy will receive a kick $v_{\rm w}$ in a random direction. We checked, however, that adopting polar outflows does not alter our results significantly (see, e.g., Fig.~\ref{fig:faint_cw}), in line with the findings of \citet{Springel2003}.  Once particles enter the wind through stochastic selection from the star-forming phase, they are briefly decoupled from hydrodynamical interactions until their SPH density falls below 10\% of the threshold density for star formation \citep[as in][]{Springel2003}. To protect against rare cases in which wind particles do not reach a region of low enough density (this could happen if, e.g., the imposed wind velocity is very low), the hydrodynamical forces are turned on again after an elapsed time equal to 2.5\% of the Hubble time at the simulation redshift.

%%%%%%%%%%%%%%%%%%%%%%%%%%%%%%%%%%%%%%%%%%%%%%%%%
\subsection{The AGN feedback model}
\label{sec:agn_model}
%%%%%%%%%%%%%%%%%%%%%%%%%%%%%%%%%%%%%%%%%%%%%%%%%

In a subset of our simulations, we include a model for the growth of supermassive black holes (BHs) and associated feedback processes. Here, we briefly summarize how these processes are followed. Full details are given elsewhere \citep{Springel2005b,Sijacki2007}. 

We assume that low-mass ``seed'' BHs are produced sufficiently frequently such that every halo above a certain threshold mass can be expected to contain at least one such BH at its centre. In the simulations, we use the on-the-fly friends-of-friends group finder that is also used for our variable wind model to put seed BHs with a mass $m_{\rm BH,seed} = 10^{4} h^{-1} M_\odot$ into halos when they exceed a mass of $5 \! \times \! 10^{10} h^{-1} M_\odot$ and do not contain any BH yet. The BHs are represented by collisionless sink particles and are allowed to grow by mergers with other BHs and by gas accretion, assuming a Bondi-Hoyle-Lyttleton prescription for the accretion rate, subject to an imposed upper limit equal to the Eddington limit. Two BHs are merged when they fall within their local SPH smoothing lengths and have sufficiently small relative velocities. At variance with previous work \citep{Springel2005b,Sijacki2007}, we do not consider any relative motion of a BH with respect to the surrounding gas when estimating its accretion rate. This was done here as the code does not follow the motion of the BHs self-consistently. Instead, they are effectively pinned to the centres of their host halos by recentering them on the potential minimum at every time step.

Motivated by growing theoretical and observational evidence that AGN feedback is composed of two characteristic modes \citep[see e.g.][]{Chrurazov2005}, we use two distinct feedback models depending on the BH accretion rate \citep[see][]{Sijacki2007}. For large accretion rates above $0.01$ of the Eddington rate, the bulk of AGN heating is assumed to be in the form of radiatively efficient quasar activity with only a small fraction of the luminosity being thermally coupled to the surrounding medium. We adopt this thermal heating efficiency to be $0.5\%$ of the rest mass-energy of the accreted gas, which reproduces the observed relation between BH mass and bulge stellar velocity dispersion \citep{DiMatteo2005}. For accretion rates below $0.01$ of the Eddington rate, we assume that feedback is in the so-called ``radio-mode'', where AGN jets inflate hot, buoyantly rising bubbles.  The mechanical feedback efficiency provided by the bubbles is taken to be $2\%$ of the accreted rest mass-energy, which is in good agreement with X--ray observations of elliptical galaxies \citep{Allen2006}. \cite{Sijacki2007} showed that this bubble model leads to a self-regulated residual BH growth in large halos and yields realistic BH mass densities. It was also found that feedback from the BHs significantly lowers the baryon fractions and X-ray luminosities of poor galaxy clusters and groups, bringing them into excellent agreement with observations \citep{Puchwein2008}. Furthermore, the luminosities of central group and cluster galaxies are reduced and made compatible with observations \citep{Puchwein2010}.

%%%%%%%%%%%%%%%%%%%%%%%%%%%%%%%%%%%%%%%%%%%%%%%%%
\subsection{Computing galaxy stellar mass functions and accounting for diffuse light}
\label{sec:get_gsmf_icl}
%%%%%%%%%%%%%%%%%%%%%%%%%%%%%%%%%%%%%%%%%%%%%%%%%

In our simulations, we find galaxies and their stellar masses by applying the {\sc Subfind} substructure finder \citep{Springel2001,Dolag2009} to FoF groups identified with a linking length of 0.2 times the mean interparticle spacing. {\sc Subfind} decomposes each FoF group into a main halo and self-bound substructures. We consider the stellar content of each of these objects to be a galaxy and compute its stellar mass by simply summing up the masses of the corresponding star particles. This allows as to compute GSMFs.

In our analysis, we also examine stellar to halo mass ratios, based  on halo central galaxies. In our simulations, the latter correspond to the star particles that are part of a FoF groups' main (sub)halo. The group's $M_{\rm 200c}$ value is adopted as halo mass and is based on a spherical overdensity mass definition\footnote{$M_{\rm 200c}$ is the mass within a spherical region around the halo centre in which the mean density is 200 times the critical density of the Universe.}.

Galaxy groups and clusters do not only contain stars that are bound to their member galaxies but also a significant diffuse stellar component \citep[e.g.,][]{Zibetti2005,McGee2010} which is also found in cosmological simulations \citep{Murante2004,Willman2004,Sommer-Larsen2005,Murante2007,Puchwein2010}. These diffuse stars are often not considered to be part of a group's central galaxy, or are simply missed altogether in shallow observations. We will, therefore, explore in Section~\ref{sec:gsmf_bright} how the massive end of our simulated GSMF changes if we exclude such diffuse stars when computing the stellar masses of central galaxies. We do this based on a surface brightness threshold. More precisely, we compute the luminosities of the star particles in the simulations in the K and V-band based on their ages and metallicities using the stellar population synthesis model library {\sc GALAXEV} \citep{Bruzual2003}. Next, we calculate the average surface brightness and enclosed stellar mass profiles of galaxies in bins of halo mass. In this way, by computing the profiles based on a large number of galaxies for each halo mass bin we avoid uncertainties in the surface brightness due to particle noise. We then derive for each halo mass bin the radius at which the profile falls below the adopted surface brightness threshold and the fraction of stellar mass within this radius. Finally, we correct the stellar masses of all galaxies in the corresponding halo mass bin by this fraction. However, in order to avoid reducing the stellar mass of low-mass galaxies which are not well resolved in the simulation, we do not remove stars that are less than one softening length from a galaxy's centre. 

We have checked that this procedure yields very similar results to computing SPH-like surface brightness estimates for all star particles, and then excluding those that fall below the threshold value and are at radii larger than one softening length. For the threshold, we adopt values commonly used in the literature as a definition of diffuse light/BCG cutoff, i.e. ${\rm V} = 26.5 \, {\rm mag} / {\rm arcsec}^2$ \citep{Rudick2006} and ${\rm K} = 20.7 \, {\rm mag} / {\rm arcsec}^2$ \citep{Lin2004}.\footnote{The BCG luminosity estimates in \citet{Lin2004} correspond to an effective cutoff at ${\rm K} \sim 20.7 \, {\rm mag} / {\rm arcsec}^2$.}

%%%%%%%%%%%%%%%%%%%%%%%%%%%%%%%%%%%%%%%%%%%%%%%%%
%%%%%%%%%%%%%%%%%%%%%%%%%%%%%%%%%%%%%%%%%%%%%%%%%
\section{Results}
\label{sec:results}
%%%%%%%%%%%%%%%%%%%%%%%%%%%%%%%%%%%%%%%%%%%%%%%%%
%%%%%%%%%%%%%%%%%%%%%%%%%%%%%%%%%%%%%%%%%%%%%%%%%

%%%%%%%%%%%%%%%%%%%%%%%%%%%%%%%%%%%%%%%%%%%%%%%%%
\subsection{The low-mass end of the galaxy stellar mass function}
\label{sec:gsmf_faint}
%%%%%%%%%%%%%%%%%%%%%%%%%%%%%%%%%%%%%%%%%%%%%%%%%

\begin{figure*}
\centerline{\includegraphics[width=\linewidth]{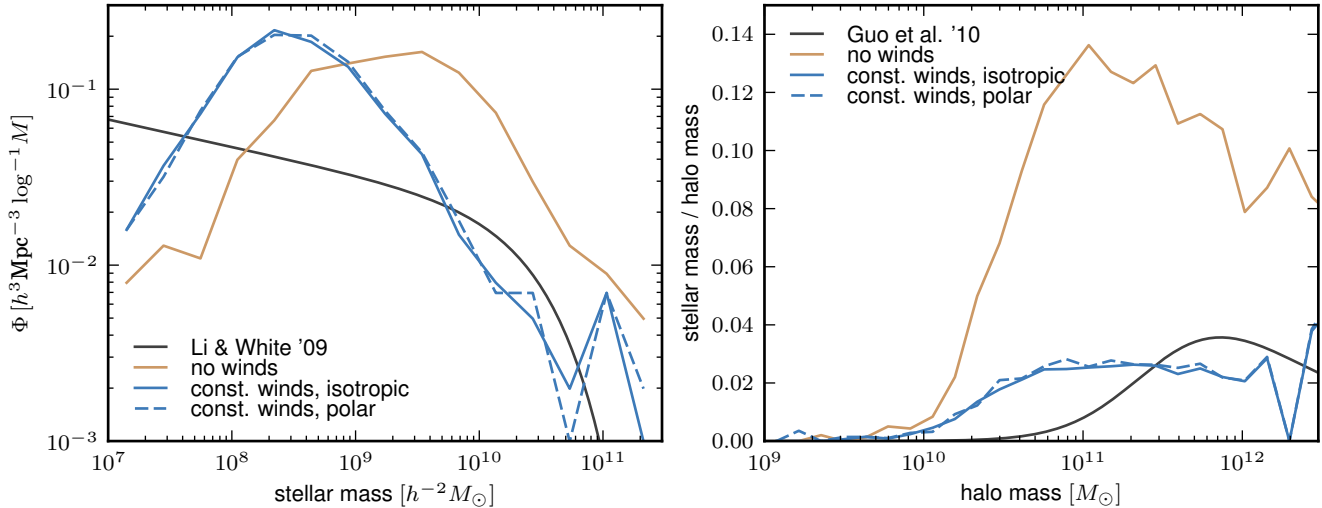}}
\caption{Impact of galactic winds with constant velocity and mass loading on the low-mass end of the GSMF ({\it left panel}) and the stellar mass fractions of halos ({\it right panel}) at $z=0$. Shown are results for a reference run without kinetic wind feedback (L15N128csf) and for simulations with isotropic (L15N128csf,cw) and polar (L15N128csf,cw,polar) winds with constant wind velocity $v_{\rm w}=\, 484 {\rm km}\, {\rm s}^{-1}$ and mass loading $\eta=2$. A fit to the observed GSMF \citep{Li2009} is shown for reference. Also shown are constraints on halo stellar fractions obtained using the abundance matching technique \citep{Guo2010}. Above the resolution limit, both runs with constant winds and without winds strongly overpredict the low-mass end of the GSMF, as well as the stellar fractions in low mass halos.}
\label{fig:faint_cw}
\end{figure*}

We first explore how the details of the wind model affect the GSMF and the stellar mass fractions of halos at redshift $z=0$. Figure~\ref{fig:faint_cw} illustrates the effects of constant winds with commonly used settings for mass loading and wind velocity ($\eta=2$ and $v_{\rm w} = 484 \, {\rm km} \, {\rm s}^{-1}$). We compare runs with isotropic and polar constant winds \citep[implemented as in][]{Springel2003} to a reference run that does not include any subresolution wind model. Observational constraints on the GSMF based on SDSS data \citep{Li2009} and halo stellar fractions obtained by the abundance matching technique \citep{Guo2010} are also shown for comparison. When winds are not included, the GSMF of well-resolved galaxies (i.e. with a stellar mass above $\sim 5 \times 10^9 h^{-2} M_\odot$) is a power-law with roughly the same steep slope as the halo mass function over the corresponding range. Accordingly, the stellar to halo mass ratio varies only mildly. It decreases slowly with increasing halo mass, but the suppression of the abundance of galaxies with stellar masses below $\sim 5 \times 10^9 h^{-2} M_\odot$ is a resolution effect (see also Fig.~\ref{fig:faint_res}).

In runs with winds, the stellar mass fractions of halos are reduced by roughly a factor of 4 to 5, somewhat less at the high-mass end, where the constant velocity winds become inefficient in expelling gas, as also found by \citet{Crain2009}. Otherwise, the suppression factor does not strongly depend on halo mass, as expected for winds that are faster than the escape velocity and have a constant mass loading. Correspondingly, the GSMF is shifted towards lower stellar masses, whereas its shape is only modified at the high-mass end. We note that the isotropic and polar wind models yield very similar results, thus the geometry of the outflows does not seem to be very important. Overall, we confirm that constant wind models are not able to reproduce the observed faint-end slope of the GSMF and fail to explain the different shapes of halo and stellar mass functions.

We thus next try variable wind models in which the wind velocity is proportional to the escape velocity of the galaxy from which the wind is launched, as suggested by observations \citep[e.g.,][]{Martin2005}. We consider both models with a momentum-driven and an energy-driven scaling of the mass loading and wind velocity. We choose the normalisation of the mass-loading such that $\eta$ is equal to 2 if the wind velocity matches the constant wind model, i.e. \begin{align}
 \eta &= 2 \times 484 \, {\rm km} \, {\rm s}^{-1} / v_{\rm w} \quad \textrm{( momentum-driven )}, \\
 \eta &= 2 \times (484 \, {\rm km} \, {\rm s}^{-1})^2 / v_{\rm w}^2 \quad 
 \left(
 \begin{array}{c}
   \textrm{energy-driven}\\
   \textrm{Salpeter IMF} 
 \end{array}
 \right).
\end{align}
For the energy-driven model, this means that the mass-loading is set by the available supernova energy, where the latter is computed assuming a Salpeter IMF \citep{Salpeter1955}. For the moment, we adopt the value $\kappa=1.25$ for the ratio of wind velocity to escape velocity. The choice of $\kappa$ will be discussed in more detail later. 

\begin{figure*}
\centerline{\includegraphics[width=\linewidth]{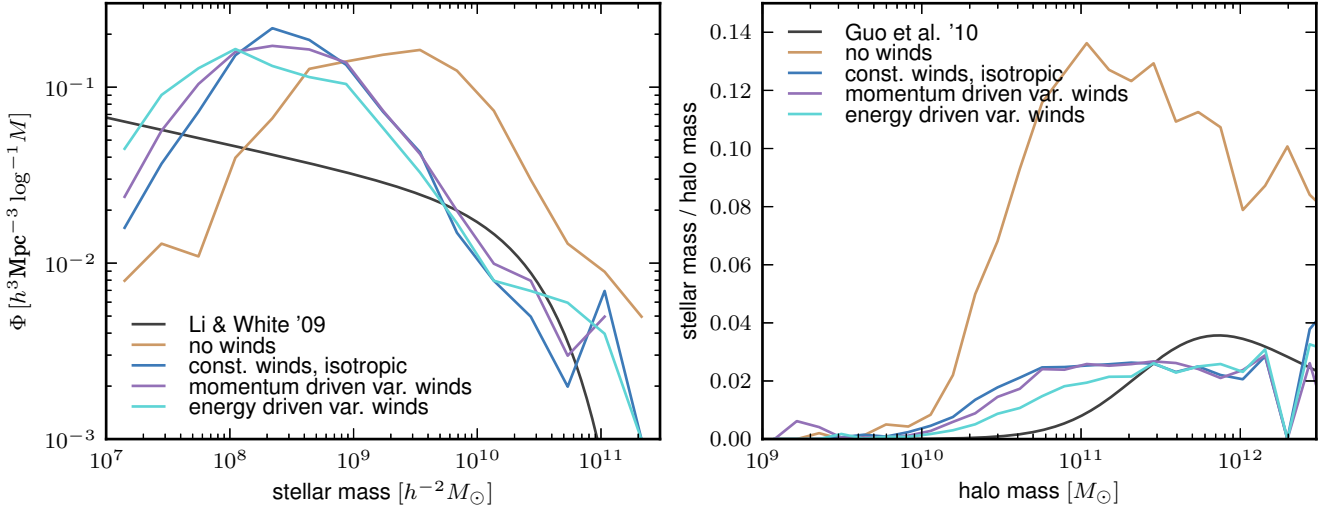}}
\caption{Impact of scaling the velocity and mass loading of galactic winds with the host halo's escape velocity on the low-mass end of the GSMF ({\it left panel}) and the stellar mass fractions of halos ({\it right panel}) at $z=0$. Shown are results for a reference run without kinetic wind feedback (L15N128csf), for simulations with isotropic winds with constant wind velocity $v_{\rm w}=\, 484\, {\rm km}\, {\rm s}^{-1}$ and mass loading $\eta=2$ (L15N128csf,cw), as well as for two variable wind models in which the wind velocity and mass loading scale with the host halo's escape velocity. In the latter models, the wind velocity is given by $v_{\rm w}=1.25\times v_{\rm esc}$. The mass loading scales with the wind velocity according to $\eta \sim v_{\rm w}^{-1}$ for the {\it momentum driven} wind model (L15N128csf,mdvw1.25) and  according to $\eta \sim v_{\rm w}^{-2}$ for the {\it energy driven} wind model (L15N128csf,edvw1.25). A fit to the observed GSMF \citep{Li2009} is shown for reference. Also shown are constraints on halo stellar fractions obtained using the abundance matching technique \citep{Guo2010}. These results illustrate that the overproduction of low mass galaxies is reduced in models in which the mass loading increases for low escape velocities.}
\label{fig:faint_scaling}
\end{figure*}

As expected and shown in Fig.~\ref{fig:faint_scaling}, a stronger scaling of mass-loading with galaxy size results in a shallower slope of the GSMF. In the energy-driven wind model, the stellar mass fractions of halos are no longer constant in the $7 \times 10^{10} - 10^{12} M_\odot$ halo mass range. Instead, the fractional stellar mass content declines for lower halo masses. This trend is also seen in the observational constraints, although with a steeper slope. We conclude from this that the energy-driven wind model appears most promising for being able to reproduce the observed shallow slope of the GSMF's low-mass end. We thus concentrate on this model and explore in the following how the choice of $\kappa$ as well as assumptions about the available energy affect its predictions.

\begin{figure*}
\centerline{\includegraphics[width=\linewidth]{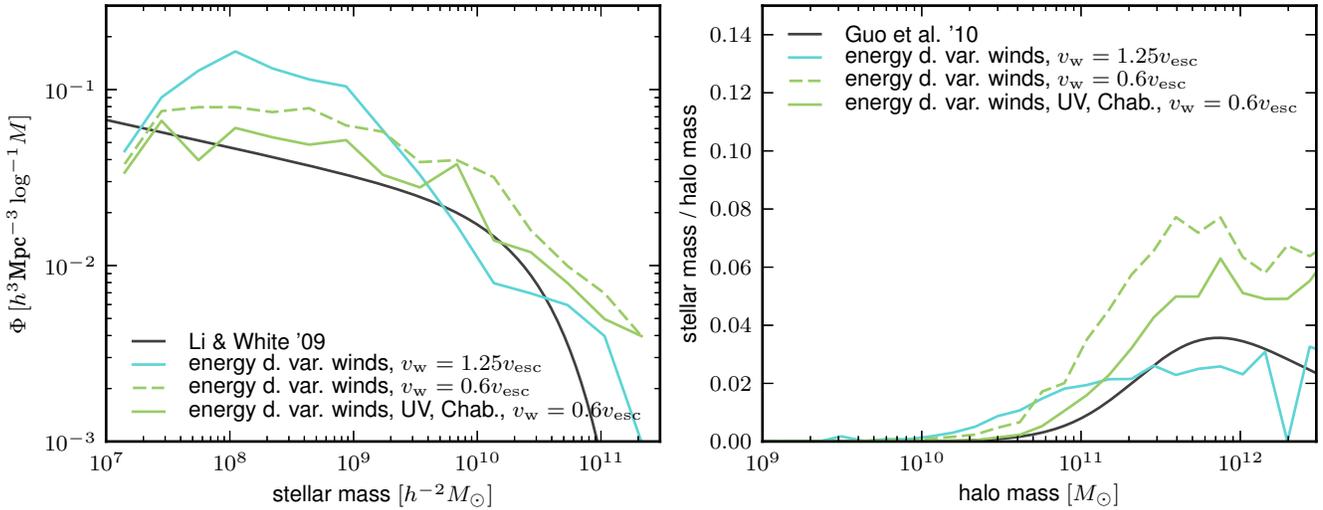}}
\caption{Dependence of the low-mass end of the GSMF ({\it left panel}) in energy driven wind models ($\eta \sim v_{\rm w}^{-2}$) on the assumed normalization of wind velocity and mass loading. The effects on the stellar mass fractions of halos are illustrated in the figure's {\it right panel}. Shown are results for $z=0$ and models in which the wind velocity is given by $v_{\rm w}=1.25\times v_{\rm esc}$ (L15N128csf,edvw1.25) and $v_{\rm w}=0.6\times v_{\rm esc}$ (L15N128csf,edvw0.6). In the latter models, winds have an accordingly larger mass loading as $\eta \sim v_{\rm w}^{-2}$. Also illustrated are the effects of increasing the available supernova feedback energy by assuming a Chabrier rather than a Salpeter IMF and of assuming more efficient photoheating ({\it solid green line}, simulation L15N128csf,edvw0.6,uv,chab, see main text for more details). A fit to the observed GSMF \citep{Li2009} is shown for reference. Also shown are constraints on halo stellar fractions obtained using the abundance matching technique \citep{Guo2010}. These results illustrate that winds with lower velocity but larger mass loading are more successful in suppressing the overproduction of low mass galaxies. As expected, boosting supernova feedback and photoheating also reduces their number.}
\label{fig:faint_vel_uv_imf}
\end{figure*}

\begin{figure*}
\centerline{\includegraphics[width=\linewidth]{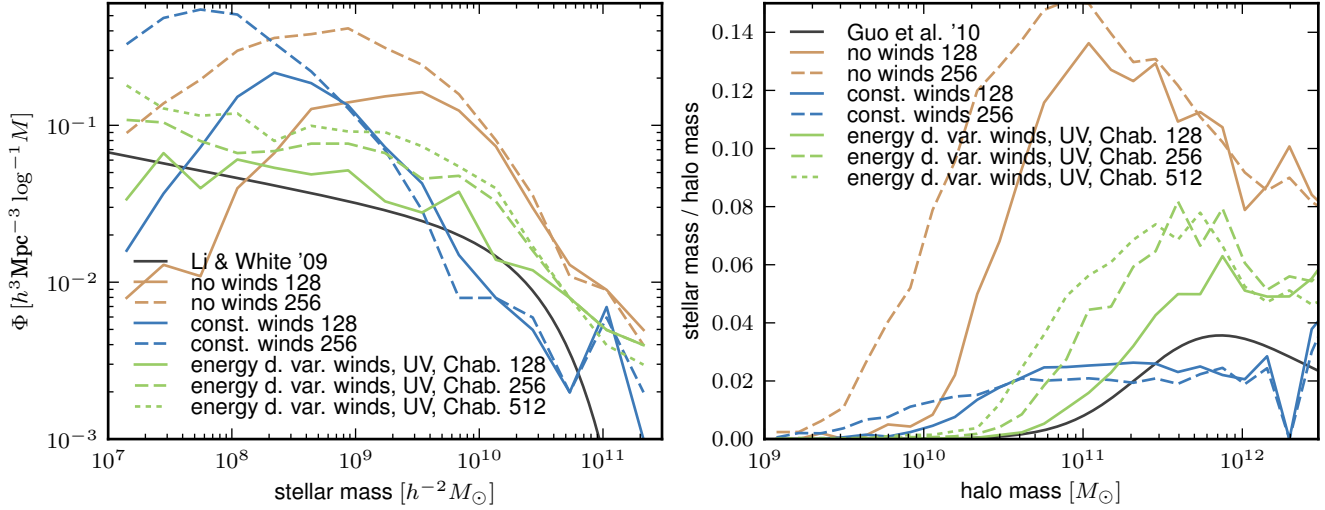}}
\caption{Dependence of the low-mass end of the GSMF ({\it left panel}) and the stellar mass fractions of halos ({\it right panel}) at $z=0$ on the simulation resolution. The effects on the stellar mass fractions of halos are illustrated in the figure's {\it right panel}. Shown are results obtained without kinetic wind feedback (L15N128csf, L15N256csf), with winds with constant velocity and mass loading (L15N128csf,cw, L15N256csf,cw), and with energy driven variable winds (L15N128csf,edvw0.6,uv,chab, L15N256csf,edvw0.6,uv,chab, L15N512csf,edvw0.6,uv,chab). All simulations shown here adopt a box size of $15\, h^{-1} \mathrm{Mpc}$. Results are given for simulations with $2\times128^3$, $2\times256^3$, and $2\times512^3$ particles ({\it solid}, {\it dashed}, and {\it dotted} lines, respectively), where the latter resolution is only available for the energy-driven variable wind model.}
\label{fig:faint_res}
\end{figure*}

Observations by \citet{Martin2005} (see their Fig.~7) found maximum wind velocities somewhat lower than the galaxies' escape velocities $v_{\rm w} \sim 0.6 \times v_{\rm esc}$. Motivated by this result we investigate whether slower energy-driven winds with $\kappa=0.6$ improve the agreement with observations. Figure~\ref{fig:faint_vel_uv_imf} shows GSMFs and stellar mass fractions for this model as well as for the previously studied $\kappa=1.25$ simulation. The figure illustrates that slower winds with a correspondingly larger mass loading are more effective in suppressing the abundance of low-mass galaxies, while the number of galaxies around the knee of the GSMF increases in such models. As a consequence, the $\kappa=0.6$ energy-driven model is able to reproduce the observed slope of the low-mass end of the GSMF. It even produces a knee at roughly the right position, however with a somewhat too shallow slope above the knee. Unfortunately, the normalisation of the predicted GSMF is still about a factor of 2 too large.

An obvious way to reduce the overall normalisation of the GSMF is to increase the amount of available feedback energy. In particular, we note that the amount of energy injected by supernovae is not accurately known, leaving room for a potential upward revision of the available energy.
Furthermore, our estimates thus far have been based on a Salpeter IMF, which implies a rather low fraction of stars that are massive enough to explode as a supernova. Using a more recent calibration of the IMF like the one by \citet{Chabrier2003} implies a factor of 1.8 larger number of supernovae per unit mass in stars formed. Adopting this revision, the mass-loading of the energy driven wind model is then given by
\begin{equation}
 \eta = 1.8 \times 2 \times (484 \, {\rm km} \, {\rm s}^{-1})^2 / v_{\rm w}^2 \quad 
 \left(
 \begin{array}{c}
   \textrm{energy-driven}\\
   \textrm{Chabrier IMF} 
 \end{array}
 \right).
\end{equation}

Another heating source that is most likely underestimated in the simulations discussed above is photoheating associated with hydrogen and helium reionization. This is implemented as in \citet{Katz1996} and \citet{Springel2003} in our current simulations. However, during a sudden reionization event, the simulation time steps in low density regions may be too large and the assumption of photoionisation equilibrium may be inaccurate \citep[see, e.g.,][for a more detailed discussion]{Peeples2010,Puchwein2012}, causing an underestimate of the energy injected during reionization.
 Furthermore, the photoheating rate is quite sensitive to the assumed spectrum of the cosmic ultraviolet background which is not well constrained. As a simple fix that allows us to get IGM temperatures that are compatible with those inferred from analyses of the Lyman-$\alpha$ forest, we try boosting the photoheating rates in the code by a factor of 3.

Both, assuming a Chabrier IMF and increasing the photoheating rate, lowers the normalisation of the GSMF and brings it into very good agreement with observations. This is illustrated in Fig.~\ref{fig:faint_vel_uv_imf}. The curves corresponding to simulations that assume a Chabrier IMF and a boosted UV heating are labelled {\it Chab} and {\it UV} in the legend of Fig.~\ref{fig:faint_vel_uv_imf}, respectively, as well as in all other figures of this paper. Checking the impact of these two changes individually, we found that increasing the supernova rate results in a larger reduction of the normalisation of the GSMF compared to boosting the photoheating rate.

To summarize, we have now found a simulation model that accurately reproduces the low-mass end of the GSMF at $z=0$. As a next step, we will explore whether this result is sensitive to changes in the simulation resolution. Figure~\ref{fig:faint_res} shows the GSMFs and stellar mass fractions for runs without winds, with constant winds, as well as for the aforementioned energy-driven variable wind model. Results are given for our standard resolution simulations and high-resolution runs with eight times better mass resolution and a two times smaller gravitational softening. For the energy-driven wind model, we also performed a very high-resolution run with 64 times better mass resolution and a four times smaller gravitational softening length.

Without winds, the results are well converged with respect to resolution above a limiting mass of $\sim 5 \times 10^9\, h^{-2} M_\odot$. Below this threshold, the GSMF in the high-resolution run keeps rising for decreasing stellar mass, while for the standard resolution it reaches a maximum and then falls off. This is a typical behaviour found in resolution studies of cosmic structure formation simulations. The runs with constant winds exhibit a very similar behaviour, albeit at lower stellar mass values.

Interestingly, the resolution limit in terms of halo mass is very similar in runs without and with constant winds, roughly lying at $\sim 7 \times 10^{10} M_\odot$, while it is, due to the much smaller stellar fractions, shifted towards lower stellar mass values in the latter runs. This suggests that accurately resolving the host halo and its potential well is more important than the number of star particles which represent a galaxy, at least for the quantities analysed here.

For the variable wind simulations, the situation is somewhat different. The slope of the low-mass end of the observed GSMF is surprisingly well reproduced at all three resolution levels even at very small stellar mass. The normalisation, however, increases somewhat between the standard and high resolution runs. Compared to the latter, the normalisation is only slightly larger in the very high resolution simulation, most likely indicating that we are approaching convergence. On second thought, it is not entirely unexpected that the convergence behaviour is different in the variable wind model. In low-mass galaxies it implies much smaller wind velocities than the constant wind scenario. Consequently, the maximum radius that is reached by an outflow is much more sensitive to the detailed structure of the galaxy, which in turn depends on the gravitational softening and the mass resolution in the simulation.

Overall, the galaxy populations in the simulations without and with constant winds seem to be well converged above the quoted resolution limits. They however fail to match the observed low-mass end of the GSMF. In the energy-driven variable wind runs, the low-mass end slope of the GSMF is reproduced well, while the normalisation is close to converged in the two higher resolution runs in the stellar mass range shown in Fig.~\ref{fig:faint_res}.    

%%%%%%%%%%%%%%%%%%%%%%%%%%%%%%%%%%%%%%%%%%%%%%%%%
\subsection{The high-mass end of the galaxy stellar mass function}
\label{sec:gsmf_bright}
%%%%%%%%%%%%%%%%%%%%%%%%%%%%%%%%%%%%%%%%%%%%%%%%%

The left panel of Fig.~\ref{fig:faint_res} suggests that supernova-driven winds do not strongly affect the abundance of large galaxies with stellar masses above $\sim 10^{11}\, h^{-1} M_\odot$. In the constant wind model the wind velocity becomes insufficient to drive gas to large radii or to expel it from these objects, while in the variable wind model the wind's mass loading becomes increasingly small for very massive galaxies.

\begin{figure*}
\centerline{\includegraphics[width=\linewidth]{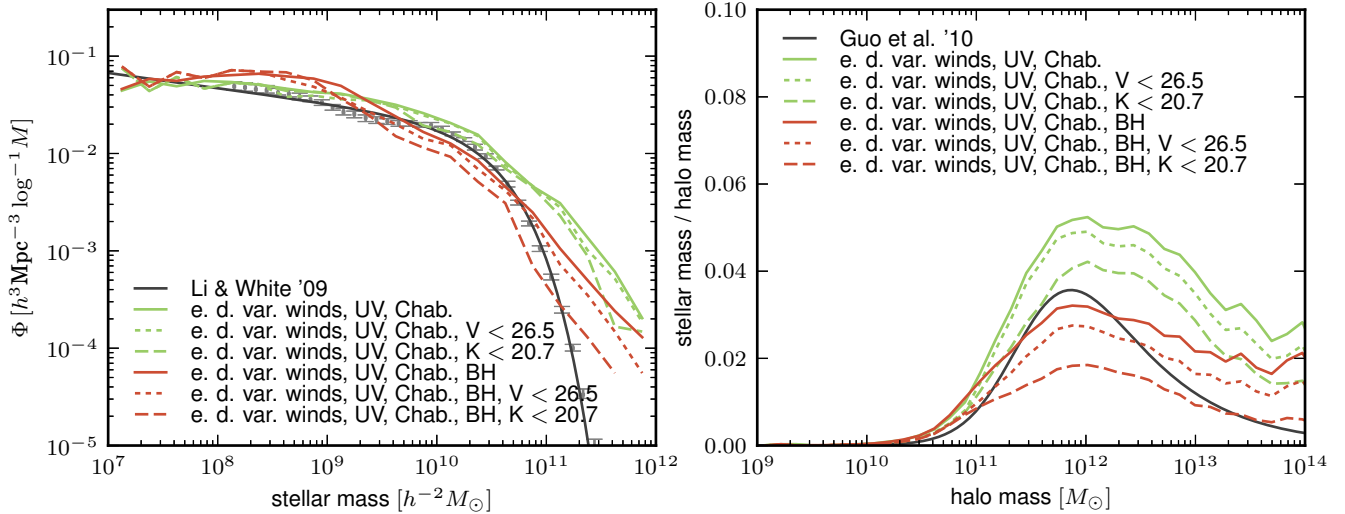}}
\caption{The impact of AGN feedback and diffuse light subtraction on the high-mass end of the GSMF ({\it left panel}) and the stellar mass fractions of halos ({\it right panel}) at $z=0$. The results are based on runs with energy driven variable winds and were performed with ({\it red curves}, L60N512csfbh,edvw0.6,uv,chab) and without black holes ({\it green curves}, L60N512csf,edvw0.6,uv,chab). We either accounted for all stars in a galaxy's host halo ({\it solid lines}) or subtracted a diffuse stellar component defined by having a V-band surface brightness dimmer than 26.5 $\mathrm{mag} \, \mathrm{arcsec}^{-2}$ ({\it dotted}) or a K-band surface brightness dimmer than 20.7 $\mathrm{mag} \, \mathrm{arcsec}^{-2}$ ({\it dashed}). Both, accounting for AGN feedback and excluding diffuse light, significantly reduces the overproduction of massive galaxies and improves the agreement with observations. The observed GSMF \citep[][{\it squares with error bars}]{Li2009} and a Schechter fit to it are shown for reference in the {\it left panel}, while the {\it right panel} displays constraints on halo stellar fractions obtained using the abundance matching technique \citep{Guo2010} for comparison.}
\label{fig:bright}
\end{figure*}

Feedback from AGN, instead, is widely considered to be a promising candidate for suppressing star formation in massive galaxies. In particular, it appears to be a necessary ingredient for explaining the high-mass end of the GSMF \citep{Croton2006, Bower2006}. Another, potentially indispensable ingredient that has been ignored in many previous numerical studies is to properly account for diffuse stellar light. A stellar component with very low surface brightness might be missed in observations or not be considered to be part of a galaxy. However, when GSMFs are measured from simulations in a straightforward way the mass of these stars is normally added to the stellar mass of the galaxy that resides at the centre of the corresponding halo. Simply because of a different accounting of the diffuse light, this can induce a discrepancy between observed and simulated GSMFs, particularly at the high-mass end, where the fraction of diffuse stars can be quite large.

Figure~\ref{fig:bright} illustrates how accounting for AGN feedback and diffuse stellar light affects the high-mass end of the GSMF. AGN feedback is included as described in Section~\ref{sec:agn_model}. A correction for diffuse stellar light is included based on a surface brightness threshold, i.e.~stars that are seen as diffuse light dimmer than 26.5 $\mathrm{mag} \, \mathrm{arcsec}^{-2}$ in the V-band or 20.7 $\mathrm{mag} \, \mathrm{arcsec}^{-2}$ in the K-band are not assigned to any galaxy and ignored when computing the GSMF. Figure~\ref{fig:bright} clearly shows that both AGN feedback and the correction for a diffuse stellar component reduce the overabundance of massive galaxies substantially, thereby helping to reproduce the high-mass end of the observed GSMF.

The AGN model employed here efficiently suppresses late-time star formation in massive galaxies \citep[see also][]{Puchwein2010}. However, its exact impact on the GSMF depends somewhat on the adopted model parameters. For the choices outlined in Section~\ref{sec:agn_model}, we get a quite good fit to the observed GSMF. We also did some tests with other parameter values and found that accurately reproducing the knee of the GSMF is sensitive to the value of the seed BH mass. More specifically, choosing a larger value for the seed black hole mass while keeping the threshold mass of the FoF groups into which they are implanted unchanged results in AGN that become too efficient in lower-mass galaxies. The abundance of galaxies around the observed knee of the GSMF is then underpredicted. Conversely, lowering the seed BH mass is accordingly expected to cause AGN to become important only for more massive galaxies, thereby shifting the knee of the simulated GSMF to larger stellar mass values. Note that the the position of the knee also depends on the exact choice of the AGN feedback parameters in semi-analytical galaxy formation models. There, typically a threshold for the ratio of cooling to free-fall time is used to control essentially the halo mass above which AGN feedback is effective, thereby shifting the knee of the GSMF \citep[e.g.][]{Bower2012}.

Correcting for a dim diffuse stellar component lowers the predicted abundance of galaxies more and more strongly for increasing stellar mass. This trend can be understood as a result of the fact that central galaxies in groups and clusters have particularly large diffuse light envelopes. Also note that we do not exclude stars from poorly resolved low-mass galaxies that are within one softening length from the galaxy's centre. As expected, the exact change in the high-mass end of the GSMF depends on the adopted surface brightness threshold value. For an accurate comparison of simulations and observations it is thus important that observational studies report the surface brightness level down to which stars are accounted for in observational constraints on the GSMF. Simulators then need to adopt this threshold in their analysis. 

To summarize, the high-mass end of the GSMF can be reproduced quite well when both AGN feedback and a correction for diffuse light is included. In combination with our energy-driven variable wind model, the predicted GSMF then matches the observations impressively well over the whole stellar mass range shown in Fig.~\ref{fig:bright}.

\begin{figure}
\centerline{\includegraphics[width=\linewidth]{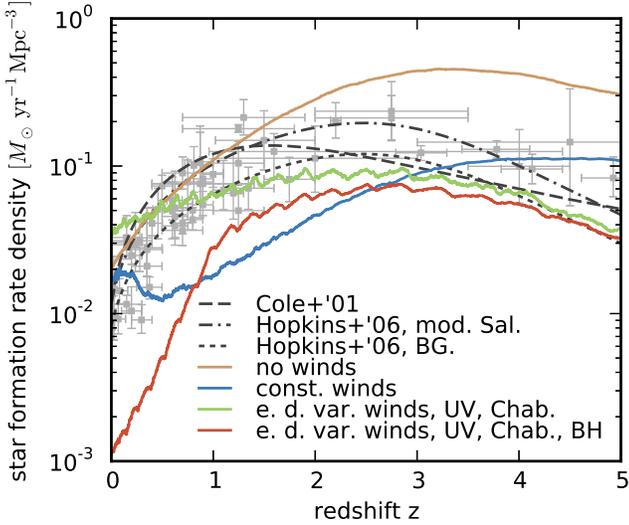}}
\caption{Star formation rate density as a function of redshift for different simulation models. Results are shown for simulations without winds (L15N256csf), as well as for runs with constant winds (L15N256csf,cw) and energy-driven variable winds (L15N256csf,edvw0.6,uv,chab). For the latter wind model also predictions based on a run with AGN feedback are shown (L60N512csfbh,edvw0.6,uv,chab). The dark grey lines indicate different fits to observational data (some of the data is also indicated by {\it squares with error bars}, see main text for more details).}
\label{fig:csfh}
\end{figure}

%%%%%%%%%%%%%%%%%%%%%%%%%%%%%%%%%%%%%%%%%%%%%%%%%
\subsection{The cosmic star formation history and high redshift stellar mass functions}
\label{sec:sfh}
%%%%%%%%%%%%%%%%%%%%%%%%%%%%%%%%%%%%%%%%%%%%%%%%%

\begin{figure*}
\centerline{\includegraphics[width=\linewidth]{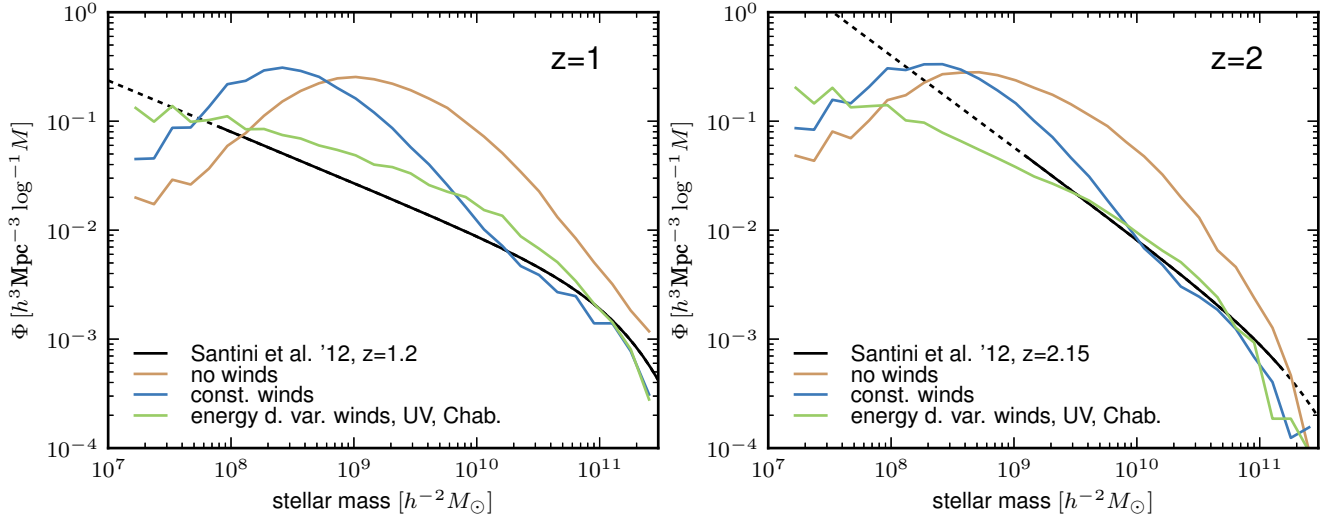}}
\caption{The impact of galactic winds on the $z=1$ ({\it left panel}) and $z=2$ ({\it right panel}) GSMF. Results are shown for runs without winds (L60N512csf), with constant winds (L60N512csf,cw), and with energy driven variable winds (L60N512csf,edvw0.6,uv,chab). Fits to the observed GSMF from \citet{Santini2012} are shown for comparison. The {\it black solid} lines indicate the stellar mass range in which the fits are constrained by the data, while the {\it black dotted} lines are extrapolations. Also at $z=1$ and 2, simulations with energy-driven variable winds are in very good agreement with the data.}
\label{fig:high_z}
\end{figure*}

In the previous sections, we have identified a feedback model that reproduces the GSMF at redshift $z=0$ well. We will now explore how the cosmic star formation rate history and high-redshift GSMFs are affected by winds and AGN, and whether the predictions of this favoured feedback model are compatible with observations.

Figure~\ref{fig:csfh} shows the star formation rate density as a function of redshift for runs without winds, as well as for simulations with constant and energy-driven variable winds. For the latter model, we also include results for a run that additionally includes AGN feedback. Fits to observational data obtained in \citet{Cole2001} and \citet{Hopkins2006} are shown for reference. In the latter case, they are displayed both for a modified Salpeter and a Baldry \& Glazebrook IMF \citep{Baldry2003}. For the results based on the Salpeter IMF also the data including their errors are indicated in the figure. The large error bars and significant scatter in the data also illustrate that the star formation rate density, especially at high redshift, is notoriously difficult to measure, in particular due to uncertainties in the IMF and the dust correction.

Energy-driven variable winds can efficiently suppress star formation at high redshift where it happens mostly in small objects, while runs without winds or with constant winds overpredict the cosmic star formation rate density for $z>1$ and $z>4$, respectively. In addition, the peak of the star formation rate density shifts to lower redshift ($z\sim2.5$) in the variable wind model, in good agreement with observations. In the other models, instead, the star formation history peaks at higher redshift, which seems disfavoured by the data. As a side note, the small ``wiggles'' in the curves of the variable wind models correspond to the times at which the on-the-fly FoF group finder is run and at which the host halo masses of all gas particles in the simulation, which are used to compute wind velocities and mass loadings, are updated.

In most of the redshift range shown in Fig.~\ref{fig:csfh}, the star formation rate density in our energy-driven variable wind simulations is slightly lower than suggested by observations. At the same time, the low-redshift GSMF seems to be slightly overpredicted. This suggests tensions between the observational constraints on the redshift evolution of the star formation rate density and the observed low-redshift GSMF as previously pointed out by, e.g., \citet{Rudnick2003} and \citet{Hopkins2006}.

At low redshift the observed star formation rate density drops almost an order of magnitude below its peak value. A similarly strong decrease can be seen in runs without winds. There it is, however, a consequence of the almost complete consumption of gas in galaxies and is closely connected to excessively large stellar mass fractions. In constant wind simulations the star formation rate density falls off too early, and is smaller than observed for $z<3$. The decline is however less steep, so that at $z=0$ it becomes comparable to the observed values again. The star formation rate density in the variable wind model also decreases at low redshift. While it remains roughly comparable to the observed values at all times, the decline seems to be somewhat less steep than observed. 

We also investigated how AGN feedback affects the cosmic star formation history and found that it does not have a huge impact at high redshift but results in a steep drop-off below $z=1$. While a strong decline at low redshift is, indeed, suggested by observations, this decline seems to start too early in our run with AGN. 

However, further test simulations showed that the onset of this drop-off is sensitive to the exact choice made for the AGN model parameters, similar to what we found for the position of the knee of the GSMF. In particular, choosing a larger seed BH mass shifts the decline to earlier times. Thus, we expect that it should also be possible to shift it to the observed later time by adopting a correspondingly tuned smaller value.

Figure~\ref{fig:high_z} displays the GSMFs at redshifts $z=1$ and 2 for runs without winds, with constant winds and with energy driven variable winds. Fits to the observed GSMF from \citet{Santini2012} are shown for comparison. These constraints are slightly offset in redshift as they were not available for the exact times of our simulation outputs. Clearly, the simulation without winds overpredicts the GSMF at $z=1$ and 2. The constant wind model is in reasonably good agreement with the $z=2$ GSMF over the limited stellar mass range in which it is constrained. At $z=1$, however, the slope of the GSMF is much steeper than observed. On the other hand, the energy-driven variable wind model is in very good agreement with the data both at redshifts $z=1$ and 2. 

Overall, the variable wind model, whose parameter choice was guided by data at $z\simeq 0$, also matches the observed cosmic star formation history and high-redshift GSMFs very well. In contrast, the constant and no wind models clearly fail in doing so.   

%%%%%%%%%%%%%%%%%%%%%%%%%%%%%%%%%%%%%%%%%%%%%%%%%
\subsection{Specific star formation rates and gas fractions of galaxies}
\label{sec:fgas}
%%%%%%%%%%%%%%%%%%%%%%%%%%%%%%%%%%%%%%%%%%%%%%%%%

Our analysis in Sections~\ref{sec:gsmf_faint} and \ref{sec:gsmf_bright} was based on the stellar masses of galaxies at $z=0$, and thus essentially probed the integral of the star formation rate form early times to the present day. In the following we will, instead, examine the instantaneous star formation rates of galaxies at $z=0$ and their dependence on stellar mass. We will also study the gas fractions of galaxies which are of course closely related to their star formation rates. 

\begin{figure}
\centerline{\includegraphics[width=\linewidth]{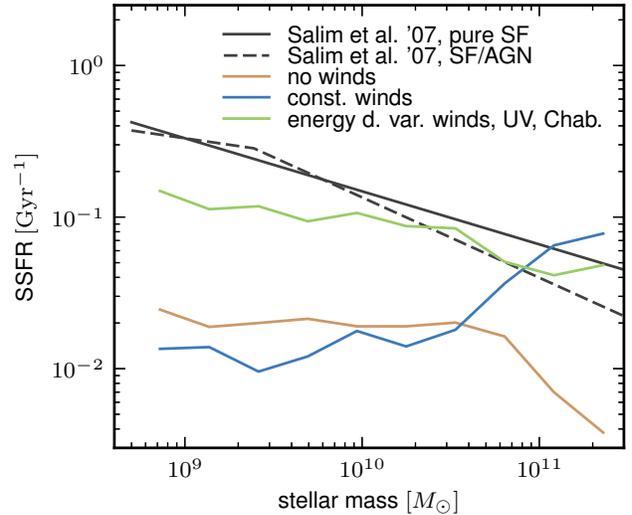}}
\caption{Specific star formation rates (SSFRs), i.e. star formation rates per unit stellar mass, for galaxies simulated without winds (L15N256csf), with constant winds (L15N256csf,cw), and with energy driven variable winds (L15N256csf,edvw0.6,uv,chab) at $z=0$. Fits to observed SSFRs from \citet{Salim2007} are displayed for reference. The {\it solid} and {\it dashed grey} lines indicate measurements for ``pure'' star-forming galaxies and star-forming galaxy/AGN composites, respectively. The energy-driven wind model produces galaxies with much larger SSFRs, thereby improving the agreement with observational constraints significantly.}
\label{fig:ssfr}
\end{figure}

Figure~\ref{fig:ssfr} shows the specific star formation rate, i.e.~the star formation rate per unit stellar mass, as a function of the stellar mass of the galaxy at $z=0$. Fits to observational data obtained by \citet{Salim2007} are shown for reference. Clearly, the constant and no wind models fail to account for the observed specific star formation rates of galaxies, while the energy-driven variable wind model is in impressively good agreement with the data.        

\begin{figure}
\centerline{\includegraphics[width=\linewidth]{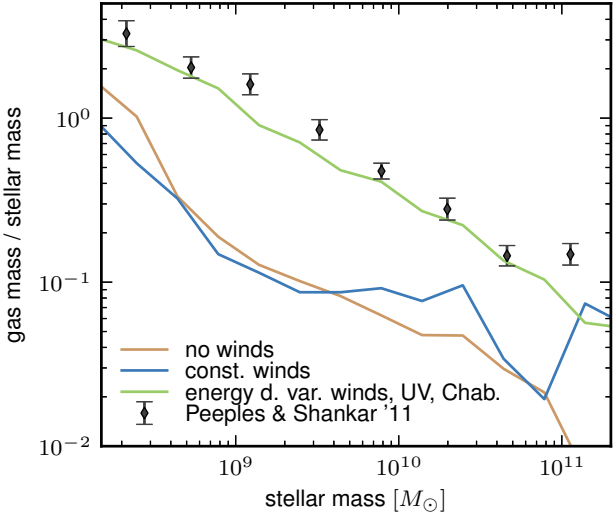}}
\caption{Mean gas mass (neutral hydrogen) to stellar mass ratios of galaxies simulated without winds (L15N256csf), with constant winds (L15N256csf,cw), and with energy driven variable winds (L15N256csf,edvw0.6,uv,chab), as a function of stellar mass at $z=0$. Binned observational data compiled by \citet{Peeples2011} are shown for comparison. The error bars indicate the standard deviations of the bins' mean values. The energy-driven variable wind model produces galaxies with much larger gas mass to stellar mass ratios, bringing them into excellent agreement with observational constraints.}
\label{fig:fgas}
\end{figure}

This can be understood by looking at the amount of cold gas in galaxies, or in other words the available fuel for star formation. Figure~\ref{fig:fgas} displays the gas mass to stellar mass ratio of galaxies as a function of their stellar mass. As we are interested in cold gas rather than in hot halo gas, and in order to make our results directly comparable to H{\sc i} observations, we included only neutral hydrogen when computing the gas masses of simulated galaxies. To this end we assumed that the star-forming multiphase interstellar medium with hydrogen densities $n_{\rm H} > 0.23 \, {\rm cm}^{-3}$ is neutral. In the employed star formation model, this is a reasonable assumption as, somewhat depending on the exact density, at least 84\% of the ISM are supposed to be in a cold phase with temperatures below $T=10^3 {\rm K}$ \citep[see][]{Springel2003}. For the remaining cold gas that falls below this density threshold and that is part of a galaxy's {\sc Subfind} (sub-)halo, the neutral hydrogen fraction is computed explicitly as a function of gas temperature and density under the assumption of collisional ionisation equilibrium in the presence of an external UV background, as in \citet{Katz1996}.

The gas mass to stellar mass ratios exhibit a similar behaviour as the specific star formation rates. Runs without and with constant winds underpredict the observed gas mass to stellar mass ratios, while the predictions based on the energy-driven variable wind model are in excellent agreement with the data. This is independent evidence that the latter model ejects just the right amount of gas from galaxies, enabling them to simultaneously reproduce both the observed stellar and gas masses. The gas fractions of the constant and no wind models look surprisingly similar. The cause for the low gas content is, however, different in these models. In the constant wind run, outflows expel a large fraction of the gas from galaxies, except in very massive objects with stellar masses above $\sim 10^{11} M_\odot$, in which the fixed wind velocity becomes insufficient to unbind gas. Without winds, the gas fractions are instead reduced by a highly efficient conversion of gas into stars. It thus appears to be a coincidence that the curves are so similar below $\sim 10^{11} M_\odot$ at $z=0$. This is also suggested by Fig.~\ref{fig:csfh} which shows that the star formation rate history is quite different in these runs, only at $z=0$ the star formation rate density happens to be remarkably similar.  

These results demonstrate that energy-driven variable winds bring not only the stellar masses and star formation histories of galaxies into much better agreement with observations but also yield gas masses and current specific star formation rates in good agreement with observational constraints. We note that this impressively good agreement with the data is found over the whole stellar mass range considered here.

%%%%%%%%%%%%%%%%%%%%%%%%%%%%%%%%%%%%%%%%%%%%%%%%%
%%%%%%%%%%%%%%%%%%%%%%%%%%%%%%%%%%%%%%%%%%%%%%%%%
\section{Summary and conclusions}
\label{sec:conclusions}
%%%%%%%%%%%%%%%%%%%%%%%%%%%%%%%%%%%%%%%%%%%%%%%%%
%%%%%%%%%%%%%%%%%%%%%%%%%%%%%%%%%%%%%%%%%%%%%%%%%

In this work, we investigated feedback processes relevant for shaping the GSMF, and how they can be accounted for in cosmological hydrodynamical simulations of galaxy formation as a subresolution model. For this purpose, we implemented a new model for galactic winds into the {\sc Gadget-3} simulation code and used it to perform a set of high-resolution simulations. This allowed us to study how wind velocity and wind mass flux need to scale with the properties of the galaxy from which the winds are launched such that the resulting simulated galaxy populations match observations. Even though identifying a favoured model in this fashion is phenomenological in nature, we emphasize that a successful wind model is a great asset for future galaxy simulations and attempts to understand the detailed physical processes that launch winds with the required properties in the first place.

We also studied the impact of feedback from AGN on the abundance of massive galaxies and assessed the importance of correctly accounting for diffuse light when comparing this quantity between simulations and observations. Our main findings are: \begin{itemize}

\item Galactic winds with constant mass-loading and velocity, which have been widely used in the literature, are unable to reproduce the observed shape and shallow faint-end slope of the GSMF.

\item Adopting a variable wind velocity equal to a specific fraction of the escape velocity of the galaxy, as suggested by observations, and allowing either for a certain amount of momentum or energy to drive the wind, much larger wind mass-loadings in low-mass galaxies are possible. This results in a shallower low-mass end of the GSMF.

\item A stronger scaling of mass-loading with wind velocity, as implied by an energy- compared to a momentum-driven wind scenario, reduces the slope of the low-mass end of the GSMF more strongly. We find that an energy-driven variable velocity wind model reproduces the observed low-mass slope well.

\item In such an energy-driven model the normalisation of the GSMF depends on the available energy for driving the wind, as well as on the adopted ratio of wind velocity to escape velocity. 

\item The observed normalisation of the low-mass-end of the GSMF can be matched well when wind velocities are adopted that are somewhat smaller than the nominal escape velocity from the centre of the halo (in agreement with observational suggestions), and when calculating the available energy for driving the wind based on the expected supernova rate.

\item Including AGN feedback and correcting for diffuse light when computing galaxy masses reduces the abundance of massive galaxies, thereby bringing the high-mass end of the GSMF into much better agreement with observations.

\end{itemize}

Overall, our simulations with energy-driven variable winds and AGN feedback reproduce the observed GSMF at redshift $z=0$ quite well. In order to gain further insights in how viable such scenarios are in light of other data, we investigated how simulations with these models compare to observational constraints on the cosmic star formation history, as well as with observations of the GSMFs at redshifts $z=1$ and 2. We found that:

\begin{itemize}

\item Energy-driven variable winds suppress star formation at high redshift, reducing the cosmic star formation rate density to observed levels and shifting its peak to $z \sim 2.5$, which is in good agreement with the data.

\item The cosmic star formation rate density declines at low redshift in the energy-driven wind model. The drop-off, however, seems to be somewhat less steep than observed, but steepens when AGN feedback is also included. 

\item When the whole redshift range of $0<z<5$ is considered, the energy-driven variable wind model reproduces the overall shape and normalisation of the observed cosmic star formation rate history much better than any of the other models.

\item The observed GSMFs at redshifts $z=1$ and 2 are reproduced very well by the energy-driven variable wind model, whereas simulations without winds fail to do so. Especially at $z=1$, runs with constant mass-loading and wind velocity also fail.

\item Over the whole stellar mass range considered, 
we find that the cold gas masses and specific star formation rates at $z=0$ are in excellent agreement with observations in runs with energy-driven variable winds.

\item In contrast, in simulations without winds or with fixed wind velocity and mass loading, the gas mass to stellar mass ratios and specific star formation rates are underpredicted by almost one order of magnitude.   

\end{itemize}

To summarise, our new energy-driven variable velocity wind model faithfully reproduces the observed low-mass end of the galaxy stellar mass function. For the first time, we combined such variable winds with a model for AGN feedback and with a proper treatment of diffuse light. Accounting for these effects allows us to simultaneously reproduce the low-mass end, the knee region, and the high-mass end of the observed GSMF.  We also obtained stellar mass fractions of halos that compare very well with results from abundance matching.

Furthermore, this model does not only make the stellar masses of galaxies compatible with observations but simultaneously almost perfectly matches their gas mass to stellar mass ratios as well as their specific star formation rates. In addition, the model drastically improves the agreement with observed GSMFs at high redshift, and with constraints on the cosmic star formation history.

This broad agreement with a wide range of data is a significant improvement compared to earlier attempts to model the buildup of the whole galaxy population in our Universe through cosmological hydrodynamical simulations. For the first time, such hydrodynamic simulations achieve a comparable success as semi-analytic models.  Simulation models that reproduce the main properties of the observed galaxy population in this fashion will be extremely helpful for making more realistic simulation predictions in the future, for example for the clustering of galaxies, for the galaxy bias, for chemical enrichment, for galaxy merger rates as a function of stellar mass, for the formation of intracluster light, etc., to just name a few.

Also, while the present work provides no insights into the detailed physics of the creation of winds in galaxies, our results strongly suggest that the mass loading in real galactic winds needs to strongly increase in low-mass galaxies, as otherwise an overproduction of the faint-end of the galaxy luminosity function seems inevitable. At the same time, our work demonstrates that the energy available from feedback processes associated with star formation is sufficient to drive such winds if losses by radiative cooling are not dominant. These findings can provide some guidance in the ongoing quest to fully understand wind formation in galaxies.

%%%%%%%%%%%%%%%%%%%%%%%%%%%%%%%%%%%%%%%%%%%%%%%%%%%%%%%%%%%%%%%%%%%%%%%
\section*{Acknowledgements}

We are grateful to Christoph Pfrommer and Tom Abel for helpful discussions. We would also like to thank Debora Sijacki and Mark Vogelsberger for carefully reading our manuscript and providing us thoughtful comments. E.P. would like to acknowledge support by the DFG through Transregio 33. V.S. acknowledges support through SFB 881, `The Milky Way System', of the DFG.

%%%%%%%%%%%%%%%%%%%%%%%%%%%%%%%%%%%%%%%%%%%%%%%%%%%%%%%%%%%%%%%%%%%%%%%
\appendix

%%%%%%%%%%%%%%%%%%%%%%%%%%%%%%%%%%%%%%%%%%%%%%%%%%%%%%%%%%%%%%%%%%%%%%%
\bibliographystyle{mn2efixed}
\bibliography{paper}
%%%%%%%%%%%%%%%%%%%%%%%%%%%%%%%%%%%%%%%%%%%%%%%%%%%%%%%%%%%%%%%%%%%%%%%
\end{document}